\documentclass[aps,twocolumn,pra,superscriptaddress,nofootinbib]{revtex4-2}
\pdfoutput=1

\usepackage{amsmath}

\usepackage{amssymb}
\usepackage{mathrsfs}
\usepackage{bm, dsfont}
\usepackage[usenames,dvipsnames]{color}
\usepackage{colortbl}
\usepackage[table]{xcolor}
\usepackage{enumitem}

\usepackage{graphicx}
\usepackage{natbib}
\usepackage[colorlinks=true,linkcolor=blue,citecolor=blue,urlcolor=blue]{hyperref}
\usepackage{cleveref}
\usepackage{hypcap}
\usepackage{verbatim, float}
\usepackage{psfrag}
\usepackage[normalem]{ulem}
\usepackage{physics}		

\newcommand{\id}{\mathds{1}}
\newcommand{\ii}{\mathrm{i}}

\newcommand{\cM}{\mathcal{M}}

\newcommand{\be}{\begin{equation}}
\newcommand{\ee}{\end{equation}}

\newcommand{\nc}{\textup{\,\o\,}}

\begin{document}

\title{Compatibility of projective measurements subject to white noise and loss}

\author{Pavel Sekatski}
\email{pavel.sekatski@unige.ch}
\affiliation{Department of Applied Physics, University of Geneva, Switzerland}

\begin{abstract}
It is well known that when two or more quantum measurements suffer from imperfections they may lose their incompatibility. For a quantum system of finite dimension $d$ we study the incompatibility of all projective measurements subjected to white noise and loss. We derive a necessary and sufficient conditions for this set of measurements to becomes compatible in terms of their visibility $p$ and efficiency $\eta$.
\end{abstract}

\maketitle

\section{Introduction}

In classical physics measurements, even the most sophisticated ones, are ontologically superfluous. Measuring a system merely reveals information on its state to an interested observer without affecting the former. All physical properties (results of possible measurements) are well defined all the time, and a super-observer could be keeping track of them while remaining unnoticed. The situation is drastically different in quantum physics. Here, several physical properties can not be simultaneously assigned to a quantum system in general. Measurements revealing such properties are usually referred to as incompatible. Then again, the "correspondence principle" suggests that in some circumstances, e.g. due to noise, coarse-graining, decoherence or whatnot, all measurements that can be performed on a quantum system are compatible and therefore appear "classical". But what are these circumstances exactly? This is the question that we address in this paper.

But first, let us recall the notion of joint measurability which provides an intuitive operational criteria formalizing the dichotomy between  (in)compatible sets of measurement,  see e.g.~\cite{HeinosaariIncompatibilityReview,GuhneJMColloquium} for reviews. We discuss it in the context of quantum theory, where a set of measurements is described by a set of positive operator valued measures $\{M_{a|x}\}$ (POVMs). Here, $x$ labels the measurement input -- it chooses the measurement from the set, and $a$ labels its output. To a state of a quantum system, modeled by a density operator $\rho$, the measurement $M_{a|x}$ associates the outcome probability distribution $\text{P}(a|x)=\tr \rho M_{a|x}$.  A set of measurements is called jointly measurable if there exists a parent POVM $\{E_\lambda\}$, such that all measurements in the set admit a decomposition 
\begin{align}\label{eq: response funcion}
M_{a|x}=\sum_\lambda p(a|x,\lambda)E_\lambda,
\end{align}
for some post-processing $p(a|x,\lambda)$. Note that, without loss of generality one can consider the post-processing to be deterministic. In words all the measurements forming a jointly measurable set are mere post-processing of a single parent measurement. The associated physical properties can be superseded by the value of $\lambda$ and assigned to the system before choosing $x$ (upon performing the parent measurement). If the construction of Eq.~\eqref{eq: response funcion} does not exist the measurement set is called not jointly measurable, or simply incompatible. In the following we will be interested in continuous-valued parent measurement, in which case one speaks of POVM density and replaces the sum with an integral in Eq.~\eqref{eq: response funcion}.

\section{Main result}

Now that we have formalized the notion of measurement compatibility, we can put our original question in a more concrete, albeit less general, form. We will consider finite-dimensional quantum systems, and study the incompatibility of all projective measurements in the circumstances where they suffer from a combination of white noise and loss -- arguably the two most common instances of measurement imperfections.  Precisely, 
the set of all projector valued measures (PVMs) in dimension $d$ is parameterized by $\{M_{a|U}= U^\dag\ketbra{a} U \}$, where $\ket{a}$  runs through the $d$ states of the computational basis with $a=0,\dots d-1$, and $U$ runs through all unitaries\footnote{When $U$ runs through all unitary transformations, $M_{a|U}= U^\dag \ketbra{a} U$ is obviously an over-parametrization of the set of all PVMs, which is totally fine for our discussion.}. To each PVM $\{M_{a|U}\}$ we associate its noisified version $\{\bar M_{a|U}^{(\eta ,p)}\}$, which suffers from white noise and loss. It is a POVM with $d+1$ outcomes given by
\be
\bar M_{a|U}^{(\eta,p)} = 
\begin{cases}
 \eta p \, M_{a|U} + \eta (1-p) \frac{\id_d}{d} & a=0,\dots,d-1  \\ 
 (1-\eta) \id_d & a= \nc
\end{cases}.
\label{eq: noisy POVM}\ee
The noisified measurement can be understood as statistical mixture of three measurements. With probability $\eta p$ it acts like the ideal PVM,  with probability $\eta(1-p)$ it produces a random outcome $a=0,\dots,d-1$, and with the remaining probability $1-\eta$ it does not "click" (these events are labeled with an additional outcome $\nc$). The two parameters are called  visibility $(p)$ and  efficiency $(\eta)$ of the measurements.

We are interested in knowing for which values of $(\eta,p)$ all PVMs in a given dimension $d$ become compatible. Note that, this questions has been addressed for two specific cases: (i) for unit efficiency $\eta=1$ in \cite{WernerState,WisemanSteering,UolaSteerJM1}, and (ii) for unit visibility $p=1$ in \cite{ioannou2022}. In the later case all lossy PVMs are incompatible for any value $\eta>0$. The following result gives a necessary and sufficient condition for the compatibility of the set of all PVMs subject to noise and loss.\\

\noindent\textbf{Result}. {\it The set of all noisified PVMs $\cM^{(\eta,p)}_d$ containing all $d$-dimensional measuremnts $\{\bar M_{a|U}^{(\eta,p)}\}$  defined in Eq.~\eqref{eq: noisy POVM} for $M_{a|U}=U^\dag \ketbra{a} U$, is jointly measurable if and only if
$ (\eta,p) \in \textbf{JM}_d$. Where the set $\textbf{JM}_d$ is closed and its boundary $\partial \textbf{JM}_d$ is given by the curve }
\be
 (\eta,p)_t = \left(T_d(t),\frac{d A_d(t) - T_d(t)}{(d-1) T_d(t) } \right).
\ee
with $t \in [0,1]$ and  
\begin{align}
A_d(t) =&   d\!\!\!\!\! \sum_{m=0}^{\min\{ \left\lfloor \frac{1}{t}-1 \right \rfloor, d-1\}} \!\binom{d-1}{m} (-1)^{d-1-m} \times \nonumber\\ \label{eq: Ad}
&\frac{ ((d-1)t(m+1) +1)(t(m+1)-1)^{d-1}}{(m+1)^2 d} \\
T_d(t)= \nonumber & d \!\!\! \sum_{m=0}^{\min\{ \left\lfloor \frac{1}{t}-1 \right \rfloor, d-1\}} \binom{d-1}{m} (-1)^{d-1-m} \\
\label{eq: Td}
&\frac{ (t(m+1)-1)^{d-1}}{(m+1)}.
\end{align}

\textit{Proof:} The full proof can be found in appendix~\ref{app: reslut 2 proof}, here we present a sketch. 

First, we observe that since the measurement assemblage $\{\bar M_{a|U}^{(\eta,p)}\}_U$  is invariant under basis change the optimal parent POVM is the covariant one. That is, the continuous-valued measurement with POVM density
\be\label{eq main: parent covariant}
E_{\bm z} = d \ketbra{\bm z}
\ee
where $\ket{\bm z} =\sum_{k=0}^{d-1} z_k \ket{k}$,   $\int \dd \bm z \ketbra{\bm z} = \frac{1}{d} \id_d$ and $\dd \bm z$ is the invariant (under unitary transformations) measure on complex vectors $\bm z\in \mathds{C}^d$ of length $|\bm z|^2=1$. 

With a fixed parent POVM, to decide if the measurement assemblage is jointly measurable it remains to check all the response functions $p(a| \bm z ,U)$ in Eq.~\eqref{eq: response funcion}. An intuitive strategy to simulate a noisified PVM in some basis $\{ U\ket{0},\dots, U\ket{d-1}\}$, is to associate each POVM element $E_{\bm z}$ to the closest state of the basis (output $k = \text{argmax}_k {|\bra{\bm z} U \ket{k}|})$, or to output the no-click outcome if all the states are far away $(\max_k {|\bra{\bm z} U \ket{k}|}<t)$. It is easy to see that this strategy simulates the POVM $N_{a|U} = U^\dag N_a U$ with
\be\label{eq: Na}
N_{a} = d \int \dd z \,\Theta_a^t(\bm z) \ketbra{\bm z},
\ee
and
\begin{align}
\Theta_k^t(\bm z)&= 
\begin{cases}
1 & |z_k|^2 \geq  \max\{ |z_0|^2,\dots, |z_{d-1}|^2, t\} \\
0 & \text{otherwise}    
\end{cases}\\
\Theta_\nc^t(\bm z)&= 
\begin{cases}
1 &  |z_0|^2,\dots, |z_{d-1}|^2< t \\
0 & \text{otherwise}    
\end{cases}.
\end{align}

Next, note that the expression in Eq.~\eqref{eq: Na} only depends on the modulus of the components $|z_k|$, and is invariant under the permutation of  values $|z_k|$ as $|z_k'|$ for $k,k'\neq a$. This symmetry implies that the simulated POVM elements are of the form
\be\label{eq: POVM simulated sym}\begin{split}
 N_k &= A_d(t) \ketbra{k} + B_d(t) \frac{\id_d-\ketbra{k}}{d-1}\\
  N_\nc &= (1-d(A_d(t)+B_d(t))) \id_d.
\end{split}
\ee
for some scalar functions $A_d(t)$ and $B_d(t)$ given by 
\be \label{eq: POVM simulated int}\begin{split}
A_d(t) + B_d(t) &=\tr N_a =   d \int \dd \bm z \Theta_0^t(\bm z) \\   
A_d(t) &= \tr N_k \ketbra{k} = d \int \dd \bm z \Theta_0^t(\bm z) |z_0|^2.
\end{split}
\ee
In the appendix \ref{app: integrals} we compute the integrals on the right hand side and show that $A_d(t)$ and $T_d(t)\equiv A_d(t) + B_d(t)$ are indeed given by Eqs.~(\ref{eq: Ad},\ref{eq: Td}). To do so we follow the approach from~\cite{lakshminarayan2008extreme}.

It remains to put the values $A_d(t)$ and $T_d(t)$ in correspondence with the noise parameters $\eta$ and $p$. Noting that for an imperfect projective measurement we have  $\tr \bar M_{a|U}^{(\eta,p)} = \eta$ and $\tr \bar{M}_{a|U}^{(\eta,p)} \ketbra{a} = \eta (p +\frac{1-p}{d})$, we conclude that our construction simulates the noisy PVMs $M_{a|U}^{(\eta,p)}$ for
\be \nonumber
\eta = A_d(t)+ B_d(t)  \quad \textrm{and} \quad p = \frac{d A_d(t) - T_d(t)}{(d-1) T_d(t) }\,.
\ee
To simulate measurements that are even more noisy, i.e. $\eta'\leq \eta$, $p'\leq p$ or both, one simply adds some noise and loss to the above construction.

Finally, to prove the "only if" direction we consider any other set of response functions $\tilde \Theta_k(\bm z)$ which lead to the same detector efficiency $d \int \tilde \Theta_k(\bm z) \dd z = \eta$ for $k=0,\dots,d-1$. Then we prove that these functions satisfy $\int \sum_{k=0}^{d-1} \tilde \Theta_k(\bm z) |z_k|^2 \dd z  \leq \int \sum_{k=0}^{d-1} \Theta_k(\bm z) |z_k|^2 \dd z $ and thus can not beat the visibility $p$ achieved with the above construction. $\square$

\section{Discussion and conclusion}
 
\begin{figure}
    \centering
    \includegraphics[width=\columnwidth]{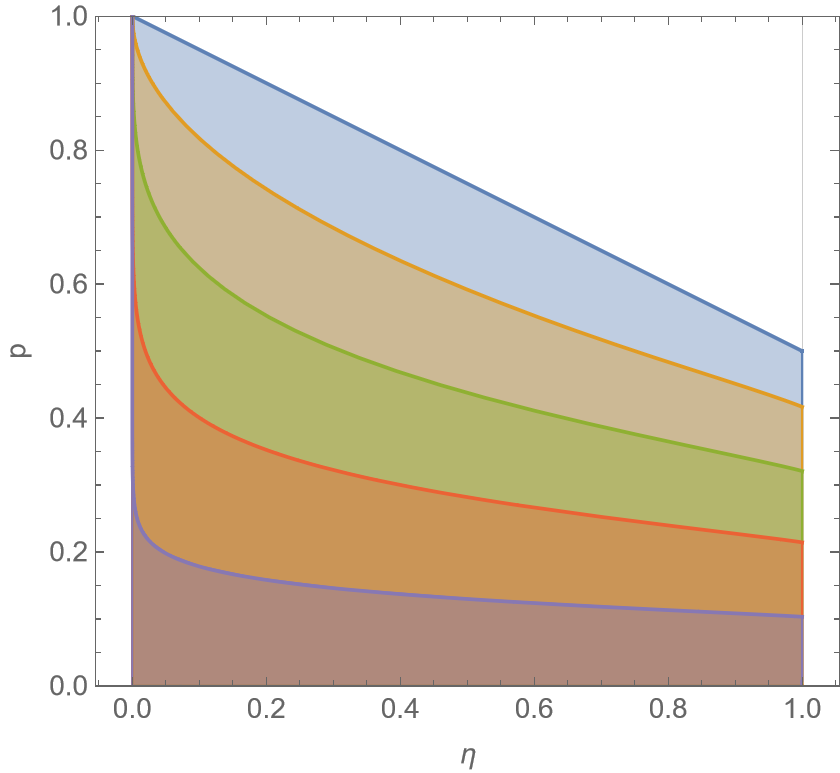}
    \caption{The set $\textbf{JM}_d$ (shaded area) for dimensions $d=2,3,5,10,30$ (from top to bottom). The set of all noisified projective measurements (PVMs) $\cM_d^{(\eta,p)}$ in dimension $d$ with visibility $p$ and efficiency $\eta$ is incompatible if and only if the point $(\eta,p)$ is above the corresponding line (outside of the shaded area).}
    \label{fig}
\end{figure}

 To illustrate the result, in Fig,~\ref{fig} we plot the set $\textbf{JM}_d$ of several value of $d$. One notes that for $d\geq 3$ the set is not convex in the $(\eta,p)$ plane. This is not in contradiction with the fact that if the measurement sets $\cM^{(\eta_1,p_1)}_d$ and $\cM_d^{(\eta_2,p_2)}$ are compatible, then the set containing their statistical mixtures $\{q \bar M_{a|U}^{(\eta_1,p_1)}+(1-q)M_{a|U}^{(\eta_2,p_2)} \}$ is also compatible. Simply the dependence of the noisified POVM on the parameters $\eta$ and $p$ is non-linear. In particular, $q M_{a|U}^{(\eta_1,p_1)}+(1-q)M_{a|U}^{(\eta_2,p_2)}= M_{a|U}^{(\eta,p)}$ with 
 \be
 \begin{split}
 \eta &= q\eta_1+(1-q)\eta_2\\
 p &=\frac{q\eta_1 p_1+(1-q)\eta_2 p_2}{q\eta_1+(1-q)\eta_2},
 \end{split}
 \ee
 and it follows from our result that $(\eta,p)\in \textbf{JM}_d$ if $(\eta_1,p_1)$ and $(\eta_2,p_2)$ are.
 
Notably, the parent measurement used in the proof to simulate the measurements sets $\cM_d^{(\eta,p)}$ is always the same -- the covariant POVM in dimension $d$. It follows, that a simple corollary of our result is that the measurement assemblage
\be
\bigcup_{(\eta,p)\in \textbf{JM}_d} \cM_d^{(\eta,p)}
\ee
containing the noisified PVMs for \emph{different} noise parameters is also jointly measurable.
 
 Next, observe that  the rather complicated functions in Eqs.~(\ref{eq: Ad},\ref{eq: Td}) become particularly simple
\begin{align}
A_d(t) &=\big(1+(d-1)t\big)  (1-t)^{d-1}    \\
T_d(t) &= d (1-t)^{d-1}
\end{align}
for large parameter values $t >\frac{1}{2}$ where $\left \lfloor \frac{1}{t}-1\right\rfloor = 0$. These parameters values correspond to $p = \frac{d A_d(t) - T_d(t)}{(d-1) T_d(t) }  = t > \frac{1}{2}$ and $\eta = T_d(t)= d(1-p)^{d-1}$. We can thus conclude that for high visibility $p>\frac{1}{2}$ the set of all PVMs is compatible if and only if
\be\label{eq: PVMs simple}
\eta \leq d(1-p)^{d-1}.
\ee

Let us now emphasize that in a recent contribution~\cite{sekatski2023unlimited}, discussing unlimited one-way steering, it was shown that the set of all noisified POVMs in dimension $d$ becomes compatible if 
\be\label{eq: POVMS}
\eta\leq (1-p)^d.
\ee 
This inequality is not necessary, and not tight with the necessary and sufficient condition for PVMs that we just derived. Indeed, one notices a factor $d$ discrepancy between the bounds of Eqs.~\eqref{eq: PVMs simple} and \eqref{eq: POVMS}. We conjecture that this is due to sub-optimality of  the POVM simulations strategy discussed in~\cite{sekatski2023unlimited}, and the main result presented here should hold for all measurements. Nevertheless, demonstrating a necessary and sufficient condition for joint measurability of all POVMs is not a en easy task, unresolved even in the simplest case of  unit efficiency $(\eta=1)$ and qubits $(d=2)$~\cite{WernerState}.

Finally, let us briefly discuss the implication of the result. In particular, what does measurement compatibility implies for the interpretation of the measurement data collected in an experiment. This depends on the context and the level of assumptions about the setup. 

On the one hand, consider the situation where the measurement model is well calibrated, i.e. one is ready to assume that the measurement is performed on a quantum system of dimension $d$ and is described by the POVM of Eq.~\eqref{eq: noisy POVM}. In this case the level of noise and loss does not matter in the limit of assymptotic sampling. Indeed, as long as $\eta,p \neq 0$ their effect on the outcome probability distribution can be inverted, by "only" paying a statistical price. 

On the other hand, in a situation where the measurement apparatus is viewed as a black box, joint measurability has very strong implications. In this case, the observed measurement data can be perfectly explained in terms of the local variables $\lambda$. In particular, this makes it impossible to observe quantum phenomena such as Einstein--Podolsky--Rosen steering and Bell nonlocality with compatible measurements and without additional assumptions on the measured states\footnote{A notable example of such assumptions is given by network scenarios, see \cite{tavakoli2022bell} for a reviewer. Here the assumption on the independence of several sources preparing the measured systems allows one to observe nonlocal correlations with all parties performing a fixed measurement (obviously compatible with itself). In this case the explanation of the observed correlations in terms of the variables $\lambda$ conflicts with the assumed network structure and is therefore untenable.}, see e.g.~\cite{UolaSteerJM1} for a more detailed discussion. 

It is worth noting that from the perspective of quantum information processing tasks with black-box measurements the necessary condition on PVM compatibility of Eq.~\eqref{eq: PVMs simple} appears rather promising. It tells us that for an arbitrary low transmission $\eta$ and a visibility $p$ above $50\%$ there is always some measurement incompatibility to exploit, provided that the dimension $d$ of the information carrier is high enough.

\emph{Acknowledgments.---} We thank Jean-Marc Luck, Serguei Sekatski, Florian Giraud, Roope Uola and Nicolas Brunner for discussions and comments. We acknowledge financial support from the Swiss National Science Foundation (projects 192244, and NCCR SwissMAP).

\bibliography{HDsteering}{}

\begin{widetext}
\appendix


\section{Extended proof}
\label{app: reslut 2 proof}

\noindent
\textbf{Result}. {\it The set of all noisy PVMs $\{\bar M_{a|U}^{(\eta,p)}\}_U$ on $\mathds{C}^d$ with $M_{a|U}=U^\dag \ketbra{a} U$ is jointly measurable if and only if
 $(\eta,p) \in \textbf{JM}_d$. Where the set $\textbf{JM}_d$ is closed and convex, and its boundary $\partial \textbf{JM}_d$ is given by the curve
\be
 (\eta,p)_t = \left(T_d(t),\frac{d A_d(t) - T_d(t)}{(d-1) T_d(t) } \right)
\ee
with $t\in [0,1]$ and }
\begin{align}
\label{eq app: Ad}
A_d(t) =&   d\!\!\!\!\! \sum_{m=0}^{\min\{ \left\lfloor \frac{1}{t}-1 \right \rfloor, d-1\}} \!\binom{d-1}{m} (-1)^{d-1-m}
\frac{ ((d-1)t(m+1) +1)(t(m+1)-1)^{d-1}}{(m+1)^2 d} \\
\label{eq app: Td}
T_d(t) = &  d \!\!\! \sum_{m=0}^{\min\{ \left\lfloor \frac{1}{t}-1 \right \rfloor, d-1\}} \binom{d-1}{m} (-1)^{d-1-m} 
\frac{ (t(m+1)-1)^{d-1}}{(m+1)}.
\end{align}\\

\textit{Proof:} First, we note that for any invariant measurement assemblage, i.e. a set   $\{M_{a|x,U}\}_{x, U}$ containing $M_{a|x,U} = U^\dag M_{a|x}U$ for all $U \in SU(d)$, the optimal parent POVM is the covariant one. 

That, is the continuous POVM with density
\be\label{eq: parent covariant}
E_{\bm z} = d \ketbra{\bm z}
\ee
with $\ket{\bm z} =\sum_{k=0}^{d-1} z_k \ket{k}$ and  $\int \dd \bm z \ketbra{\bm z} = \frac{1}{d} \id_d$, where $\dd \bm z$ is the uniform measure on complex vectors $\bm z\in \mathds{C}^d$ of length $|\bm z|^2=1$. Remark, that the covariant POVM is usually defined with the density $E_\psi =d\ketbra{\psi}$ where $\psi$ runs through all pure \textit{quantum states}. Albeit the output $\bm z$ in $E_{\bm z}$ contains an irrelevant global phase that does not affect the POVM element $\ketbra{\bm z}=\ketbra{\psi}$, these definition are equivalent for the purposes of this paper.

By optimality of the covariant POVM we mean here that if the assemblage $\{M_{a|x}\}$ is jointly measurable by some parent POVM, it is also jointly measurable by $\{E_{\bm z}\}$. A formal proof of this statement can be found in Result 2 of~\cite{ioannou2022} in a more general case of optimal compression instruments. The intuition behind is however quite simple -- if some parent POVM $E_\lambda$ does the job, its rotated version $U^\dag E_\lambda U$ does it too. Furthermore, so does the "randomly rotated" parent POVM -- which becomes equivalent to the covariant POVM when the random unitary $U$ is sampled from the Haar measure on $U(d)$.

Since the optimal parent POVM is known, to decide if an invariant measurement assemblage $\{\bar M^{(\eta,p)}_{a|U}\}$ is jointly measurable it remains to check all the response functions $p(a| \bm z ,U)$ in Eq.~\eqref{eq: response funcion}. We now define a particular response function and then show that it is the optimal one.

To simulate the PVM in the basis $\{ U\ket{0},\dots, U\ket{d-1}\}$ we use the deterministic response function
\be
a(\bm z, U) =
\begin{cases}
\text{argmax}_k |\bra{\bm z}U\ket{k}|^2 & \max_k |\bra{\bm z} U\ket{k}|^2 \geq t \\
\nc & \text{otherwise}.
\end{cases}
\ee

The measurements resulting from the covariant parent POVM in Eq.~\eqref{eq: parent covariant} followed by this response function $a(\bm z, U)$ are
\be
N_{a'|U} = d \int \dd \bm z\,  \delta_{a',a(\bm z, U)} \ketbra{\bm z}.
\ee
By symmetry we have $N_{a'|U} = U^\dag N_{a'|\id_d} U$ and  will only consider the case $N_{a'}\equiv N_{a'|\id_d}$ in the following. Defining the functions 
\be \label{app : theta det}
\Theta_k^t(\bm z)= 
\begin{cases}
1 & |z_k|^2 \geq  \max\{ |z_0|^2,\dots, |z_{d-1}|^2, t\} \\
0 & \text{otherwise}
\end{cases}
\ee
for $k=0,\dots,d-1$ and 
\be\label{app : theta det 2}
\Theta_{\nc}^t(\bm z) = 
\begin{cases}
1 & t >  \max\{ |z_0|^2,\dots, |z_{d-1}|^2\} \\
0 & \text{otherwise}
\end{cases},
\ee
one easily sees that
\be
N_a = d \int \dd \bm z \, \Theta_a^t(\bm z) \ketbra{\bm z}. 
\ee
Let us briefly study the symmetries of this operator focusing on $N_0$. First, we note that it is invariant under phase transformations, that is for $U_{\bm\varphi} = \sum_{k=0}^{d-1} e^{\ii \varphi_k} \ketbra{k}$ we have
\be\label{app: symetri argument}\begin{split}
U_{\bm \varphi} N_0 U_{\bm \varphi}^\dag &= d \int \dd \bm z \, \Theta_a^t(\bm z) U_{\bm \varphi}\ketbra{\bm z} U_{\bm \varphi}^\dag \\
    &= d \int \dd \bm z \, \Theta_a^t(\bm z) \ketbra{\bm z'} \\
    &= d \int \dd \bm z' \, \Theta_a^t(\bm z') \ketbra{\bm z'}\\
    & = N_0
\end{split}
\ee
where we used the fact that both $\Theta_a^t(\bm z)$  and the integration measure are by definition invariant under transformations that only change the phases of the complex vector components $z'_k = z_k e^{\ii \varphi_k}$. This symmetry implies that $N_0$ is diagonal in the computational basis
\be
N_0 = U_{\bm \varphi} N_0 U_{\bm \varphi}^\dag = \int \dd \bm \varphi U_{\bm \varphi} N_0 U_{\bm \varphi}^\dag = \sum n_k \ketbra{k}.
\ee
Next we note that $N_0= U_\pi N_0 U_\pi^\dag$ is invariant under unitary transformations $U_\pi$ that permute the states of the computational basis $U_\pi \ket{k} = \ket{\pi(k)}$ but leave the state $\ket{0}$ untouched ($\pi(0)=0$). This is again a consequence of the fact that the function $\Theta_a^t(\bm z)$ and the integration measure are unaffected by such variable change, which can be used analogously to Eq.~\eqref{app: symetri argument}. This symmetry implies that 
\be
N_0 = U_{\pi} N_0 U_{\pi}^\dag = \frac{1}{\sum_{\pi|\pi(0)=0}}\sum_{\pi|\pi(0)=0}  U_{\pi} N_0 U_{\pi}^\dag = n \ketbra{0} + n' \sum_{k=1}^{d-1} \ketbra{k}.
\ee
The same logic applies for all $N_k$, which can thus be expressed as
\be
 N_k = A_d(t) \ketbra{k} + B_d(t) \frac{\id_d-\ketbra{k}}{d-1},
\ee
where $A_d(t)$ and $B_d(t)$ are some scalar function.
Finally, to get the remaining POVM element $N_\nc$ note that the functions $\Theta^t_{a}(\bm z)$ for $a=0,\dots,d-1,\nc$ do not overlap (except on a set of measure zero) and cover the whole complex sphere, hence 
\be
  N_\nc = \id_d -\sum_{k=0}^{d-1} N_k = \big(1-d A_d(t)- d B_d(t)\big) \id_d.
\ee

It remains to compute the values of the functions
\be \begin{split}
T_d(t)\equiv A_d(t) + B_d(t) &=\tr N_a =   d \int \dd \bm z\, \Theta_0^t (\bm z) \\   
A_d(t) &= \tr N_k \ketbra{k} = d \int \dd \bm z\, \Theta_0^t (\bm z) |z_0|^2.
\end{split}
\ee
Performing these integrals is a little technical, and their computation is presented in a separate section\ref{app: integrals}. There we show that $A_d(t)$ and $T_d(t)$ defined above are indeed given by Eqs.~(\ref{eq app: Ad}, \ref{eq app: Td}). Here we will continue with the proof of the Result.

Noting that for the noisified PVM in the computational basis we have  
\be
\tr \bar M_a^{(\eta,p)} = \eta \qquad \text{and} \qquad \tr \bar{M}_a^{(\eta,p)} \ketbra{a} = \eta (p +\frac{1-p}{d}),
\ee
we can identify the parameters $\eta$ and $p$ with the values  $T_d(t) =\eta$ and $A_d(t)=\eta (p +\frac{1-p}{d})$. We can thus conclude that our construction simulates the imperfect PVMs with
\be\label{eq app: curve}\begin{split}
\eta(t) &= A_d(t)+ B_d(t)\\
p(t) &= \frac{d A_d(t) - T_d(t)}{(d-1) T_d(t) }.
\end{split}
\ee
To simulate any PVMs $\bar M_a^{(\eta',p')}$ that is even more noisy, i.e. $\eta'\leq \eta$, $p'\leq p$ or both, one simply adds some noise and loss to the above construction. In the $(\eta, p)$-plane the curve $(\eta(t), p(t))$ of Eq.~\eqref{eq app: curve} is the boundary of the (closed) region where all imperfect PVMs are jointly measurable with the strategy presented above.\\

Let us now show that outside this parameter region all imperfect PVMs are incompatible. We can show this by proving that the response function we used is the optimal one (we already know that the choice of parent POVM is optimal). To do so consider any other (not necessarily deterministic) response functions $\tilde \Theta_k(\bm z)$ which simulates the imperfect PVM in the computational basis for detector efficiency $\eta$ and visibility $\tilde p$. Such functions define a post-processing of the parent POVM, i.e. they associate an outcome probability distribution $ \left(\tilde \Theta_0(\bm z),\dots,\tilde \Theta_{d-1}(\bm z), \tilde \Theta_\nc(\bm z)\right)$ to each point $\bm z$ on the complex sphere. As such they must be positive $\tilde \Theta_a(\bm z) \geq 0$ and satisfy 
\be \label{app: tilde norm}
\sum_{k=0}^{d-1} \tilde \Theta_k(\bm z) + \tilde \Theta_\nc(\bm z) =1
\ee
for all $\bm z$ except a set of measure zero. By construction these response functions simulate the POVM with elements 
\be
\tilde N_a = \int \tilde \Theta_a(\bm z) E_{\bm z} \dd \bm z = d \int \tilde \Theta_a(\bm z) \ketbra{\bm z} \dd \bm z.
\ee
Since by assumption $\{N_a\}$ simulates the imperfect PVM with efficiency $\eta$ we find
\be\begin{split}
d \int \tilde \Theta_k(\bm z) \dd z &=  \tr \tilde N_k =\eta \qquad \text{for} \qquad k=0,\dots,d-1.\\
d \int \tilde \Theta_k(\bm z) \dd z &= \tr \tilde N_\nc =d(1-\eta).
\end{split}
\ee

Let us now consider the quantity $\sum_{k=0}^{d-1} \int \tilde \Theta_k(\bm z) |z_k|^2 \dd z$. On the one hand it is directly related to the visibility $\tilde p$ of the simulated mesurements 
\be
\sum_{k=0}^{d-1} \int \tilde \Theta_k(\bm z) |z_k|^2 \dd z =  \sum_{k=0}^{d-1}   \tr \tilde N_k \ketbra{k} = d \, \eta \left( \tilde  p + \frac{1-\tilde p}{d}\right).
\ee
On the other, it satisfies 
\be
\begin{split}
\sum_{k=0}^{d-1} \int \tilde \Theta_k(\bm z) |z_k|^2 \dd z &\leq  
\sum_{k=0}^{d-1} \int \tilde \Theta_k(\bm z) \max_j |z_j|^2 \dd \bm z \\
&= \int \sum_{k=0}^{d-1} \tilde \Theta_k(\bm z) \max_j |z_j|^2 \dd \bm z \\
&=\int \left(1- \tilde \Theta_\nc(\bm z)\right) \max_j |z_j|^2 \dd z.
\end{split}
\ee
But 
\be
\int  \tilde \Theta_\nc(\bm z) \max_j |z_j|^2 \dd z \geq \int \Theta_\nc(\bm z) \max_j |z_j|^2,
\ee
because by construction, among all functions with area $\int  \tilde \Theta_\nc(\bm z)  \dd z =1-\eta$ the function $\Theta_\nc(\bm z)$ collects the vectors $\bm z$ with the minimal $\max_j |z_j|^2$. Therefore, we get
\be
 d \, \eta \left( \tilde  p + \frac{1-\tilde p}{d}\right) = \sum_{k=0}^{d-1} \int \tilde \Theta_k(\bm z) |z_k|^2 \dd z \leq \int \left(1- \Theta_\nc(\bm z) \right) \max_j |z_j|^2 \dd z = d \, \eta \left(   p + \frac{1- p}{d}\right)
\ee
implying that $p\geq \tilde p$. Hence for a given efficiency $\eta$ it is impossible to simulate imperfect PVMs with a visibility $p$ higher than achieved by deterministic response functions $\Theta_a^t(\bm z)$ in Eqs.~(\ref{app : theta det},\ref{app : theta det 2}) (for the value of $t$ corresponding to $\eta$), proving the "only if" direction. Note that here we didn't restrict the consideration to deterministic response functions, so this also includes any statistical mixture of simulation strategies (response functions).  To see that the set $\textbf{JM}_d$ is closed simply note that it contains its boundary.  $\square$

\subsection{Solving the integrals on the complex sphere}

\label{app: integrals}

The purpose of this appendix is to complete the proof  by solving the integrals defining the functions $A_d(t)$ and  $T_d(t)$. Recall that these integrals are 
\be
T_d(t) =   d \int \dd \bm z\, \Theta_0^t (\bm z) \,\,\,\,\qquad A_d(t) =d \int \dd \bm z\, \Theta_0^t (\bm z) |z_0|^2 
\ee
where 
\be
\Theta_0^t(\bm z) = \begin{cases}
1 & |z_0|^2 \geq  \max\{ |z_0|^2,\dots, |z_{d-1}|^2, t\} \\
0 & \text{otherwise}
\end{cases}.
\ee

Here $\dd \bm z$ is the invariant measure on the complex $d$-sphere, i.e. the set of vectors $\bm z =(z_0,\dots,z_{d-1}) \in \mathds{C}^d$ with unit length $|\bm z|=1$. More precisely  that the measure is invariant under unitary transformations of the vectors (corresponding to basis changes in the Hilbert space). The simplest way to express this uniform measure on the complex sphere is by taking the flat measure $\Pi_{i=0}^{d-1} \dd z_i' \dd z_i''$ on $\mathds{C}^d$ , where $z_i'$ ($z_i''$) is the  real (imaginary) part, and multiplying it by a delta function $\delta(|\bm z|^2-1)$. Note that the flat measure is manifestly invariant under unitary coordinate transformations, since the Jacobian is precisely given by the unitary matrix and thus has determinant of unit modulus. Therefore, our integral over the complex sphere can be expressed as 
\be\label{app: invariant masure}
\int \dd \bm z   =
\frac{(d-1)!}{(\pi)^d} \int \Pi_{i=0}^{d-1} \dd z_i' \dd z_i'' \,  \delta(|\bm z|^2-1).
\ee
As an exercise one can verify that the normalization $\frac{(d-1)!}{(\pi)^d} \int \Pi_{i=0}^{d-1} \dd z_i' \dd z_i'' \delta(|\bm z|^2-1) =1$ is correct, though it will also follow from the following calculations. 

We are interested in the following quantities
\be
I_d(t|\text{f})= \int \dd \bm z\,  \Theta_0^t(\bm z) \text{f}( |z_0|^2 )
\ee
Using Eq.~\eqref{app: invariant masure} and changing the variables to $z_i = \sqrt{s_i} e^{\ii \theta_i}$, with $\theta_i \in [0,2\pi]$, $s_i \in [0,\infty)$ and $\dd z_i'\dd z_i '' = \frac{1}{2} \dd \theta_i \dd s_i$, we obtain
\be\begin{split}
I_d(t|\text{f})&= \frac{(d-1)!}{(\pi)^d} \int \Pi_{i=0}^{d-1} \dd z_i' \dd z_i'' \,  \Theta_0^t(\bm z) \text{f}( |z_0|^2 ) \delta(|\bm z|^2-1) \\
&= \frac{(d-1)!}{(2 \pi)^d} \int_{0}^{2\pi} \dd \theta_0 \dots \dd \theta_{d-1}
\int_0^{\infty} \dd s_0 \dots \dd s_{d-1} \Theta_0^t(\bm s) \text{f}(s_0)  \delta(\sum_i s_i -1) \\
& = (d-1)!
\int_0^{\infty} \dd s_0 \dots \dd s_{d-1} \Theta_0^t(\bm s) \text{f}(s_0)  \delta(\sum_i s_i -1)
\end{split}
\ee
Here we slightly abused the notation using $\Theta_0^t(\bm s) = \begin{cases}
1 & s_0 \geq  \max\{ s_0,\dots, s_{d-1}, t\} \\
0 & \text{otherwise}
\end{cases}$. Note that $\Theta_0^t(\bm s)$ is an indicator function which verifies that $s_0 \geq t$ and all the other coordinates are smaller $s_i\leq s_0$. Hence, it can be absorbed into the integration boundaries to give 
\be
I_d(t|\text{f}) =  (d-1)!
\int_t^{\infty} \dd s_0 \int_{0}^{s_0} \dd s_1 \dots \dd s_{d-1} \text{f}(s_0)  \delta(\sum_i s_i -1).
\ee

Next, following \cite{lakshminarayan2008extreme}, where similar integrals are computed, we use the integral representation of the delta function  $\delta(x)=\frac{1}{2\pi} \int_{-\infty}^\infty \dd \xi e^{\ii \xi x} $. In addition, for convenience we displace the integration contour from the real line and write
\be
\delta(x) = \lim_{\varepsilon \to 0_+} \frac{1}{2\pi} \int_{-\infty}^\infty \dd \xi e^{\ii (\xi+\ii \varepsilon) x }.
\ee
We will now first compute the expression
\be
    I_d^{(\varepsilon)}(t|\text{f})=\frac{(d-1)!}{2\pi} \int_{-\infty}^\infty \dd \xi \int_0^{\infty} \dd s_0 \dots \dd s_{d-1} \Theta_0^t(\bm s) \text{f}(s_0)  e^{\ii (\xi+\ii \varepsilon) (\sum_i s_i -1) },
\ee
and then take the limit $I_d(t|\text{f}) = \lim_{\varepsilon \to 0_+} I_d^{(\varepsilon)}(t|\text{f}) $. We have 

\be\begin{split}
I_d^{(\varepsilon)}(t|\text{f})& = \frac{(d-1)!}{2\pi}
\int_t^{\infty} \dd s_0 \, \text{f}(s_0) \int_0^{s_0} \dd s_1 \dots \dd s_{d-1} 
\int_{-\infty}^\infty \dd \xi e^{\ii (\xi+\ii \varepsilon) (\sum_i s_i -1)} 
\\
&=  \frac{(d-1)!}{2\pi}\int_{-\infty}^\infty \dd \xi e^{-\ii (\xi+\ii \varepsilon)}
\int_t^{\infty} \dd s_0 \, \text{f}(s_0) e^{\ii s_0 (\xi+\ii \varepsilon)} \left(\int_0^{s_0} \dd s_1 e^{\ii s_1 (\xi+\ii \varepsilon) }\right)^{d-1} 
\\
&= \frac{(d-1)!}{2\pi}\int_{-\infty}^\infty \dd \xi e^{-\ii (\xi+\ii \varepsilon)}
\int_t^{\infty} \dd s_0 \, \text{f}(s_0) e^{\ii s_0 (\xi+\ii \varepsilon)}   \left(\frac{e^{\ii s_0 (\xi+\ii \varepsilon)}-1}{\ii (\xi+\ii \varepsilon)}\right)^{d-1} 
\\
&=  \frac{(d-1)!}{2\pi}\int_{-\infty}^\infty \dd \xi \frac{e^{-\ii (\xi+\ii \varepsilon)}}{(\ii (\xi+\ii \varepsilon))^{d-1}}
\int_t^{\infty} \dd s_0 \,  \text{f}(s_0) e^{\ii s_0 (\xi+\ii \varepsilon)} \left(e^{\ii s_0 (\xi+\ii \varepsilon)}-1\right)^{d-1} 
\\
&=  \frac{(d-1)!}{2\pi}\int_{-\infty}^\infty \dd \xi  \frac{e^{-\ii (\xi+\ii \varepsilon)}}{(\ii (\xi+\ii \varepsilon))^{d-1}} \sum_{m=0}^{d-1}\binom{d-1}{m} (-1)^{d-1-m} \int_t^\infty \dd s \, \text{f}(s) e^{\ii s (m+1)(\xi+\ii \varepsilon)}.
\end{split}
\label{eq app: sum repres}
\ee
Below we compute this expression for the two cases of interest, $\text{f}_0(s)=1$ and $\text{f}_1(s)=s$, using Cauchy's residue theorem.\\

\paragraph{Zero's moment}
Let us start with the zero's moment $\text{f}_0(s)=1$. We have
\be
\int_t^\infty \dd s \, \text{f}_0(s) e^{\ii s (m+1)(\xi+\ii \varepsilon)}
  = \left[\frac{e^{\ii s(m+1)(\xi+\ii \varepsilon)}}{\ii (\xi+\ii \varepsilon) (m+1)}\right]_{t}^{\infty} = - \frac{e^{\ii t(m+1)(\xi+\ii \varepsilon)}}{\ii (\xi+\ii \varepsilon) (m+1)}.
\ee
It remains to perform the  integration with respect to $\dd \xi$, omitting the combinatorial factors one finds
\be
\frac{1}{2\pi}   \int_{-\infty}^\infty \dd \xi  \frac{e^{-\ii (\xi+\ii \varepsilon)}}{(\ii (\xi+\ii \varepsilon))^{d-1}}  \frac{e^{\ii t(m+1)(\xi+\ii \varepsilon)}}{\ii (\xi+\ii \varepsilon) (m+1)} = \frac{1}{2\pi (m+1)} \int_{-\infty}^\infty \dd \xi  \frac{e^{\ii (t(m+1)-1)(\xi+\ii \varepsilon)}}{(\ii (\xi+\ii \varepsilon))^{d}}.
\ee
To take the integral we close the contour by adding a half-circle at infinity and use Cauchy's residue theorem. Depending on the sign of $t(m+1)-1$ the contour is closed differently.\\

For $g = t(m+1)-1 > 0$ the contour is closed with a positive imaginary part, that is $\xi = R e^{\ii \theta}$ with $\theta \in [0,\pi]$ and $\sin(\theta)\geq 0$. Then 
$R \left|\frac{e^{\ii g \xi}}{\xi^d}\right| = R \frac{e^{- g R \sin(\theta) }}{R^d} \to 0 $
for $R\to \infty$. The function $\frac{e^{\ii (x(m+1)-1)(\xi+\ii \varepsilon)}}{(\ii (\xi+\ii \varepsilon))^{d}} $ has no singularities in the complex half-plane with positive imaginary part. The integral is thus zero.\\  

In contrast, for $g = t(m+1)-1 < 0$ the contour is closed with a negative imaginary part, that is $\xi = R e^{\ii \theta}$ with $\theta \in [-\pi,0]$ and $\sin(\theta)\leq 0$. Such that 
$ R \left|\frac{e^{\ii g \xi}}{\xi^d}\right| = R \frac{e^{ g R \sin(\theta) }}{R^d} \to 0$
for $R\to \infty$. The function $\frac{e^{\ii (x(m+1)-1)(\xi+\ii \varepsilon)}}{(\ii (\xi+\ii \varepsilon))^{d}} $ has a singularity in the complex half-plane with negative imaginary part at $\xi = -\ii \varepsilon$. The integral is then computed with the residue theorem (the contour is negatively oriented) to give
\be\begin{split}
 \frac{-1}{2\pi (m+1)} \int_{-\infty}^\infty \dd \xi  \frac{e^{\ii (t(m+1)-1)(\xi+\ii \varepsilon)}}{(\ii (\xi+\ii \varepsilon))^{d}} &= \frac{1}{2\pi (m+1)} \frac{2\pi \ii}{(d-1)!} \lim_{\xi \to -\ii \varepsilon} \frac{\dd^{d-1}}{\dd \xi^{d-1}}\frac{ 
 e^{\ii (t(m+1)-1)(\xi+\ii \varepsilon)}}{\ii^d} \\
 & =\frac{1}{2\pi (m+1)} \frac{2\pi \ii}{(d-1)!}\frac{ 
 (\ii (t(m+1)-1))^{d-1}}{\ii^d} \\
 & =\frac{ (t(m+1)-1)^{d-1}}{(m+1)(d-1)!}.
\end{split}
\ee

Now note that the condition $t(m+1)-1< 0$ is always fulfilled for $m=0$ (since we only care of $t<1$). Furthermore, it can be expressed as $m< \frac{1}{t}-1$. The sum over integer $m$ thus runs until $m = \left\lfloor \frac{1}{t}-1 \right \rfloor$, and we find
\be \label{app: zero moment}
I_d(t|\text{f}_0) = \sum_{m=0}^{\min\{ \left\lfloor \frac{1}{t}-1 \right \rfloor, d-1\}} \binom{d-1}{m} (-1)^{d-1-m} \frac{ (t(m+1)-1)^{d-1}}{(m+1)},
\ee
which proves Eq.~\eqref{eq app: Td}.
A particularly simple situation is $t=0$, where the sum runs through all terms and one finds
\be
I_d(0|\text{f}_0) = \sum_{m=0}^{d-1} \binom{d-1}{m} (-1)^{d-1-m} \frac{ (-1)^{d-1}}{(m+1)} = \frac{1}{d} \sum_{m=0}^{d-1} \binom{d}{m+1} (-1)^m =\frac{1-(1-1)^d}{d}= \frac{1}{d}.
\ee
Which is the probability that $|z_0|^2$ is bigger that all the other components. It also verifies that the normalization of the measure was correct.\\

\paragraph{First moment}

We now compute $I_d(0|\text{f}_1)$ for the first moment $\text{f}_1(s)=s$. Define
\be
\begin{split}
    K_m(\xi+\ii\varepsilon,t) & =\int_t^\infty \dd s \, \text{f}_1(s) e^{\ii s (m+1)(\xi+\ii \varepsilon)} \\
    & = \left[ \frac{e^{\ii s(m+1)(\xi+\ii \varepsilon)}(\ii s (m+1)(\xi +\ii \varepsilon)-1)}{(\ii (\xi+\ii \varepsilon) (m+1))^2} \right ]_{t}^{\infty} \\
    & = - \frac{e^{\ii t(m+1)(\xi+\ii \varepsilon)}(\ii t (m+1)(\xi +\ii \varepsilon)-1)}{(\ii (\xi+\ii \varepsilon) (m+1))^2}      
\end{split}
\ee
Using Cauchy's residue theorem in the same fashion as above we find for $t(m+1)-1<0$
\be\begin{split}
\frac{1}{2\pi} \int_{-\infty}^{\infty} \dd \xi \frac{e^{-\ii (\xi+\ii \varepsilon)}}{(\ii (\xi+\ii \varepsilon))^{d-1}} K_m(\xi+\ii \varepsilon,t) &=
\frac{\ii}{(m+1)^2 d!}\lim_{\xi \to -\ii \varepsilon}\frac{\dd^d }{\dd \xi^d} \frac{e^{\ii (t(m+1)-1)(\xi+\ii \varepsilon)}(\ii t (m+1)(\xi +\ii \varepsilon)-1)}{\ii^{d+1}}\\
&=
\frac{\ii}{(m+1)^2 d!} \frac{d \ii t (m+1) (\ii (t(m+1)-1))^{d-1}- (\ii (t(m+1)-1))^d}{\ii^{d+1}}\\
& =\frac{ ((d-1) t(m+1) +1)(t(m+1)-1)^{d-1}}{(m+1)^2 d!}.
\end{split}
\ee
Plugging this expression in the sum of Eq.~\eqref{eq app: sum repres} gives
\be \label{app: first moment}
I_d(0|\text{f}_1) = \sum_{m=0}^{\min\{ \left\lfloor \frac{1}{t}-1 \right \rfloor, d-1\}} \binom{d-1}{m} (-1)^{d-1-m} \frac{ ((d-1) t(m+1) +1)(t(m+1)-1)^{d-1}}{(m+1)^2 d},
\ee
proving Eq.~\eqref{eq app: Ad}. As before the case of $t=0$ is particularly simple and yields
\be\label{eq app: last}\begin{split}
I_d(0|\text{f}_1) &= \sum_{m=0}^{d-1} \binom{d-1}{m} (-1)^{d-1-m} \frac{(-1)^{d-1}}{(m+1)^2 d} = \frac{1}{d^2}\sum_{m=0}^{d-1} \binom{d}{m+1}  \frac{(-1)^{m}}{m+1} 
\\
&= -\frac{1}{d^2}\sum_{m=1}^{d} \binom{d}{m}  \frac{(-1)^{m}}{m} 
= \frac{H_d}{d^2},  
\end{split}
\ee
where 
$H_d = \sum_{k=1}^d \frac{1}{k}$ is the harmonic number. 
To obtain the last equality consider the following integral
equal to the harmonic number \be
\int_0^1 \frac{1-t^d}{1-t}\dd t = \int_0^1 \frac{(1-t)\sum_{k=0}^{d-1} t^k }{1-t}\dd t  = \int_0^1 \sum_{k=0}^{d-1} t^k \dd t    = \sum_{m=0}^{d-1} \frac{1}{k+1} = H_d.
\ee
By performing a variable change $t=1-s$ we see that it is  also equal to the expression appearing before the last equality in Eq.~\eqref{eq app: last}.
\be\begin{split}
\int_0^1 \frac{1-t^d}{1-t}\dd t &= \int_0^1 \frac{1-(1-s)^d}{s}\dd s = 
\int_0^1 \frac{1}{s}\left(1-\sum_{m=0}^{d} \binom{d}{m} (-s)^m\right)\dd s 
= 
  - \int_0^1 \frac{1}{s}\sum_{m=1}^{d} \binom{d}{m} (-s)^m\dd s  \\
  &= -  \sum_{m=1}^{d} (-1)^m \binom{d}{m} \int_0^1  s^{m-1}\dd s = -  \sum_{m=1}^{d} \binom{d}{m} \frac{(-1)^m}{m}.
\end{split}
\ee
\end{widetext}
\end{document}